\documentstyle[preprint,prc,aps]{revtex}
\begin{document}
\catcode`\@=11
\def\eqalign#1{\null\,\vcenter{\openup\jot\m@th
\ialign{\strut\hfil$\displaystyle{##}$&$\displaystyle{{}##}$\hfil
     \crcr#1\crcr}}\,}
\catcode`\@=12
\title{HEAVY-FLAVOUR PENTAQUARKS IN A CHIRAL CONSTITUENT QUARK MODEL}
\author{M. Genovese\footnote{Supported by the EU program ERBFMBICT 950427}
 and J.-M. Richard}
\address{Institut des Sciences Nucl\'eaires, Universit\'e Joseph Fourier-IN2P3
-CNRS, 53, avenue des Martyrs, F-38026 Grenoble Cedex, France} 
\author{Fl. Stancu}
\address{Universit\'{e} de Li\`ege, Institut de Physique B.5, Sart Tilman,
B-4000 Li\`ege 1, Belgium}
\author{S. Pepin}
\address{Theoretical Physics Group, Department of Physics and Astronomy,
 University of Manchester, Manchester M13 9PL, United Kingdom}
\date{\today}
\maketitle
\begin{abstract}
Within the chiral constituent quark model of Glozman and Riska, we discuss the
stability of heavy pentaquarks, i.e. hadrons containing four light 
quarks and a heavy 
antiquark. 
The spin-dependent part of the Hamiltonian is dominated by the 
short-range part of the Goldstone-boson-exchange interaction. We find that 
these systems are not bound, having an energy above the lowest dissociation 
threshold into a baryon and a meson. 
\end{abstract}

\vspace{1cm}

The question of the existence of hadrons other than mesons $q\bar{q}$ and 
baryons $q^3$ remains open. Multiquark states as tetraquarks $q^2\bar{q}^2$,
 pentaquarks $q^4\bar{q}$ or hexaquarks $q^6$, could exist in QCD. Based
on a model with a chromomagnetic interaction, Jaffe \cite{Ja77} predicted the
existence of several multiquark states and among them the $uuddss$ hexaquark,
with $J^P=0^+$ and $I=0$, known as the $H$-particle. 
Since then, 
production and decay properties of such exotic hadrons have been derived 
within several models and extensive efforts have been made to find them 
experimentally \cite{Man95}.
The recent high-sensitivity search at Brookhaven \cite{St97} gives
no evidence for the production of deeply bound $H$-particles, as predicted by
Jaffe.

This is the present situation in the light sector. However, from theoretical 
general arguments \cite{JMR94,MW93,Zou86}, one expects an increase in the 
stability of the multiquark system containing a heavy-flavour quark $Q$. 
Experiments are being planned at 
Fermilab and CERN to search for new heavy hadrons and in particular for doubly
charmed tetraquarks \cite{Mo96,Ka95,Ba96}. Very recently, a search 
for the pentaquark 
$P_{\bar{c}s}^0 = \bar{c}suud$ and $P_{\bar{c}s}^- = \bar{c}sddu$ production,
 performed at Fermilab, has been reported \cite{Ai97}. Decay properties 
into the channel
$\phi\pi p$ have been analysed assuming a life time ranging from 0.1 to 1 ps 
and a mass between 2.75 and 2.91 GeV.

This is an incentive to study the stability under strong decays of 
pentaquarks $q^4\bar{Q}$ where $q=u$, $d$ or $s$ and $Q=c$ or $b$. 
From a theoretical point of view, the field is about ten years old. In the 
constituent quark model based on one-gluon exchange (OGE) interaction between
quarks, the existence of a stable compact pentaquark was first predicted in
Refs. \cite{Gi87,Li87} provided that at least one of the light quarks is
strange and 
$Q=c$ or $b$. But with a proper treatment of the kinetic energy and a realistic
breaking of the flavour $SU_F(3)$ symmetry, the pentaquark
turned out to be unstable \cite{Fl89,ZR94}. A systematic study of all 
pentaquarks has been performed 
in Ref.\cite{LSB89}
where over a dozen are found to be candidates for stability. Among them 
those with strangeness
$S=-1$ or $-2$ were the most favourable. Calculations using an instanton 
\cite{TNK93} or a Skyrme model \cite{RS93} indicate that pentaquarks, which
are not necessarily strange, can appear as bound states or as 
 near-threshold resonances, depending on the parameters of the model. On the 
other hand, the possible existence of heavy-flavour pentaquarks as 
molecular-type resonances, where the quarks interact by the long-range 
one-pion exchange 
interaction, has also been considered \cite{Sh95}. In those calculations, only
the $N\bar{B^*}$ system was found to be bound. Finally, a study of charmed
multiquarks in a bag model, where no bound state is found,
has also recently appeared \cite{charmlets}.

The present study can be viewed as a natural extension of our previous
investigations \cite{PSGR97} on the stability of heavy-flavour tetraquarks 
within the chiral constituent quark model of Glozman and Riska 
\cite{GR96,GPP96,semrel,GR96+}. In this model, the
hyperfine splitting in hadrons is due to the short-range part of the Goldstone
boson exchange (GBE) interaction, instead of the one-gluon exchange (OGE) used
in conventional models. In Ref.\cite{PSGR97}, we found that the GBE interaction
strongly binds the heavy tetraquark system $QQ\bar{q}\bar{q}$ where $Q=c$ or 
$b$ ; this is at variance with the models based on OGE interaction where 
$cc\bar{q}\bar{q}$ is unstable. Here we also find results
somehow different from the previous literature. We show that the
GBE interaction induces a short-range repulsion, of several hundreds of
MeV in both $P_{\bar{c}}$ and $P_{\bar{b}}$ systems. However, the GBE
interaction also generates a long-range attraction due to its Yukawa-type
potential. Together with a correlated two pseudoscalar-meson exchange, this
will induce a long- and a medium-range attraction which could favour a
deuteron-size (molecular type) bound state. In the following, we restrict
ourselves to the study of compact objects.

In the spirit of Jaffe's pioneering work \cite{Ja77}, here we raise the
question whether or not the model of Glozman and Riska can accomodate a
strongly bound, compact pentaquark. We shall compare our results with
those based on OGE, as mentioned above. 

The model of Glozman and Riska reproduces well the light-quark baryon
spectrum 
\cite{GR96,GPP96,semrel} and some extension to heavy baryons
has also been proposed \cite{GR96+}; in particular, it gives
the correct ordering of positive and negative parity states in all parts of
the considered spectrum. 
The detailed form of the Hamiltonian used here will be given below. First it
is useful to consider a simplified GBE interaction of the form (where radial
dependence is neglected):
\begin{equation}
 V_{\chi} = - C_{\chi} \sum_{i<j}  \lambda_{i}^{F} . \lambda_{j}^{F}
\vec{\sigma}_i . \vec{\sigma}_j ,
\label{opFS}
\end{equation}
where $\lambda_{i}^{F}$ ($F=1,2,...,8$) are the quark-flavour Gell-Mann
matrices (with an implicit summation over $F$) and $\vec{\sigma}$ are the spin
matrices. The minus sign of the interaction (\ref{opFS}) is related to the
sign of the short-range part of the GBE interaction, crucial for the
hyperfine splitting in baryon spectroscopy. This feature of the short-range
part of the GBE interaction is discussed at length by Glozman and
Riska \cite{GR96}. A typical order of magnitude
for the constant $C_{\chi}$ is about 30 MeV.

Here, we study pentaquarks with strangeness $S=0,-1,-2$ and $-3$. As in
Ref.\cite{PSGR97}, we neglect the interaction due to meson exchange
between a light quark and a heavy antiquark. Then the GBE interaction in a 
pentaquark $q^4\bar{Q}$ reduces to the interaction between the light 
quarks $q=u,d$ or $s$.

The subsystem of four light quarks must be in a colour {\bf 3} state, or 
alternatively $[211]_C$, in order to give rise to a colourless pentaquark. To
calculate the GBE matrix elements, we
assume that all light quarks are identical and that the ground-state orbital 
wave function is symmetric under permutation of light quarks. The
corresponding Young diagram notation is $[4]_O$. Then all 
combined spin $[f]_S$ and
flavour $[f]_F$ symmetries, allowed by Pauli principle, form the states 
indicated in the first column of Table \ref{states}. They have been obtained from inner
product rules \cite{book} applied to the permutation group $S_4$. In 
column 2, the corresponding
values of the spin $S$ are indicated for each $[f]_S$ representation of 
$SU_S(2)$. The total angular momentum of the pentaquark is $\vec{J}= \vec{S}
+ \vec{s}_Q$ where $s_Q=1/2$. The isospin $I$ associated to a given $SU_F(3)$
representation is given in columns 3--6 for four distinct contents of the
four-quark subsystem with strangeness $S=0, -1, -2$ and $-3$. The last column
indicates the expectation value of the operator (\ref{opFS}) which can be
conveniently calculated from the formula:
\begin{equation}
\langle V_{\chi} \rangle = N (N-10) + \frac{4}{3} S(S+1) + 2 C_F + 4 C_C
\end{equation}
where $N=4$, $C_C=4/3$ and $C_F$ is the $SU_F(3)$ Casimir operator eigenvalue.
One can see that the states in column 1 are given in an increasing order of
$\langle V_{\chi} \rangle$. Thus $|[22]_S [211]_F \rangle$ is most 
favourable (negative) for the 
structure $uuds\bar{Q}$ ($I=1/2$, $J=0$ or $1$) or $udss\bar{Q}$ ($I=0$, $J=0$ 
or $1$), $|[31]_S [22]_F \rangle$ for $uudd\bar{Q}$ ($I=0$, $J=1/2$ or $3/2$) 
and $|[22]_S [31]_F \rangle$
for $usss\bar{Q}$ ($I=1/2$, $J=1/2$). The column
$\langle V_{\chi} \rangle$ also suggests that
other $q^4\bar{Q}$ systems, with $IJ$
different from the above values, would be heavier, thus there
is no reason to present results for such obviously unstable states. 

The GBE Hamiltonian has the form \cite{GPP96} :
\begin{equation}
H= \sum_i m_i + \sum_i \frac{\vec{p}_{i}^{\,2}}{2m_i} - \frac {(\sum_i
\vec{p}_{i})^2}{2\sum_i m_i} + \sum_{i<j} V_{\text{conf}}(r_{ij}) + \sum_{i<j}
V_\chi(r_{ij}) \, ,
\label{ham}
\end{equation}
with the linear confining interaction :
\begin{equation}
 V_{\text{conf}}(r_{ij}) = -\frac{3}{8}\lambda_{i}^{c}\cdot\lambda_{j}^{c} \, C
\, r_{ij} \, ,
\label{conf}
\end{equation}
and the spin--spin component of the GBE interaction in its $SU_F(3)$ form :
\begin{eqnarray}
V_\chi(r_{ij})
&=&
\left\{\sum_{F=1}^3 V_{\pi}(r_{ij}) \lambda_i^F \lambda_j^F \right.
\nonumber \\
&+& \left. \sum_{F=4}^7 V_{K}(r_{ij}) \lambda_i^F \lambda_j^F
+V_{\eta}(r_{ij}) \lambda_i^8 \lambda_j^8
+V_{\eta^{\prime}}(r_{ij}) \lambda_i^0 \lambda_j^0\right\}
\vec\sigma_i\cdot\vec\sigma_j,
\label{VCHI}
\end{eqnarray}
 
\noindent
with $\lambda^0 = \sqrt{2/3}~{\bf 1}$, where $\bf 1$ is the $3\times3$ unit
matrix. The interaction (\ref{VCHI}) contains $\gamma = \pi, K, \eta$ or 
$\eta'$ meson-exchanges and the form of $V_{\gamma}(r_{ij})$ is given 
explicitly in
Ref.\cite{GPP96} as the sum of two distinct contributions: a Yukawa-type
potential containing the mass of the exchanged meson and a short-range
contribution of opposite sign, the role of which is crucial in baryon
spectroscopy. For a given meson $\gamma$, the meson exchange potential is :
\begin{equation}V_\gamma (r)=
\frac{g_\gamma^2}{4\pi}\frac{1}{12m_i m_j}
\{\Theta(r-r_0)\mu_\gamma^2\frac{e^{-\mu_\gamma r}}{ r}- \frac {4}{\sqrt {\pi}}
\alpha^3 \exp(-\alpha^2(r-r_0)^2)\}
,  \hspace{5mm} \gamma = \pi, K, \eta, \eta' . 
\label{POINT} \end{equation}

For the Hamiltonian (\ref{ham})--(\ref{POINT}), we use the
parameters of Ref.\cite{GPP96}. These are :
$$\frac{g_{\pi q}^2}{4\pi} = \frac{g_{\eta q}^2}{4\pi} = 
\frac{g_{Kq}^2}{4\pi}= 0.67,\,\,
\frac{g_{\eta ' q}^2}{4\pi} = 1.206 , $$
$$r_0 = 0.43 \, {\rm fm}, ~\alpha = 2.91 \, {\rm fm}^{-1},~~
 C= 0.474 \, {\rm fm}^{-2}, \, m_{u,d} = 340 \, {\rm MeV}, $$
\begin{equation}
 \mu_{\pi} = 139 \, {\rm MeV},~ \mu_{\eta} = 547 \, {\rm MeV},~
\mu_{\eta'} = 958 \, {\rm MeV},~ \mu_{K} = 495 \, {\rm MeV}.
\label{PAR} \end{equation}
They provide a very satisfactory description of low-lying nonstrange baryons,
extended to strange baryons in Ref.\cite{semrel} where a dynamical
three-body calculation is performed as well. The latter reference gives $m_s=0.440$ GeV.

First, we calculate 
variationally the masses of the threshold hadrons using the Hamiltonian
(\ref{ham})--(\ref{PAR}). The meson mass is obtained
from a trial wave function of type $\Phi \propto \exp{[-\alpha r^2/2]}$ with
$\vec{r}=\vec{r}_1-\vec{r}_2$ by varying the parameter $\alpha$. The
heavy-quark mass is adjusted to reproduce the experimental average mass
$\bar{M}=(M+3 M^*)/4$ of $M=D$ or $B$ mesons. This gives $m_c = 1.35$ GeV and
$m_b = 4.66$ GeV. With all quark masses now fixed, we estimate 
the baryon masses with a trial wave function of type $\Phi \propto 
\exp{[-\alpha
 (\rho^2 + \lambda^2)/2]}$ where $\vec{\rho} = \vec{r}_2-\vec{r}_3$ and 
$\vec{\lambda} = (2 \vec{r}_1 - \vec{r}_2 - \vec{r}_3)/\sqrt{3}$. The results
for baryons and mesons are exhibited in Table \ref{hadron}, together with the
experimental values. The theoretical values lie at most $\sim 50$ MeV above
the experiment. The baryons masses, obtained from the same Hamiltonian
but in a more precise Fadeev type 
calculations \cite{semrel} are also indicated. 

For the pentaquarks, it is useful to consider the following system of internal 
Jacobi coordinates:
\begin{equation}
\begin{array}{ccl}
\vec{x} &=& \vec{r}_1 - \vec{r}_2 ,\\
\vec{y} &=& (\vec{r}_1 + \vec{r}_2 - 2 \vec{r}_3)/\sqrt{3} ,\\
\vec{z} &=& (\vec{r}_1 + \vec{r}_2 + \vec{r}_3 - 3\vec{r}_4)/\sqrt{6} ,\\
\vec{t} &=& (\vec{r}_1 + \vec{r}_2 + \vec{r}_3 + \vec{r}_4 - 4 
\vec{r}_5)/\sqrt{10} .
\end{array}
\end{equation}
 Then we assume a variational wave function of the form:
\begin{equation}
\Psi = \left(\frac{a}{\pi}\right)^{9/4} \left(\frac{b}{\pi}\right)^{3/4} 
\exp{[-\frac{a}{2}(x^2+y^2 + z^2) - \frac{b}{2} t^2]}.
\label{wf}
\end{equation}
 The expectation value of the kinetic energy T involves
 the average of the inverse light masses defined as:
\begin{equation}
4/\mu_1 = \left\{ \begin{array}{c} 1/m_a + 3/m_b \hspace{5mm} \mbox{for} 
\hspace{5mm} q_aq_{b}^{3} ,\\ 
 2/m_a + 2/m_b \hspace{5mm} \mbox{for} \hspace{5mm} q_{a}^{2}q_{b}^{2} ,
\end{array} \right.
\end{equation}
for the light subsystem. With
\begin{equation}
\frac{1}{\mu_2} = \frac{1}{4\mu_1} + \frac{1}{m_Q} ,
\end{equation}
the internal 
kinetic energy of the $q^4\bar{Q}$ system reads
\begin{equation}
T = \frac{3}{2} \left(\frac{3}{\mu_1}a + \frac{4}{5\mu_2} b\right).
\end{equation}
For the calculation of the confinement potential energy, we first need to 
evaluate the colour operator matrix elements :
\begin{equation}
\langle O^C \rangle = \langle\sum_{i<j}^{4} \frac{\lambda^{c}_{i}}{2} . 
\frac{\lambda^{c}_{j}}{2} \rangle
= 1/2(C_C - 4 C_q) , 
\label{colour}
\end{equation}
where $C_q = 4/3$. For the colour state $[211]_C$ one has $C_C = 4/3$. Hence
$<O^C>_{[211]} = -2$.  As there are 6 pairs in (\ref{colour}) one gets
\begin{equation}
\langle\frac{\lambda^{c}_{i}}{2} . \frac{\lambda^{c}_{j}}{2}\rangle 
= -1/3 , \hspace{2cm} i,j = q ,
\end{equation}
for each pair. For the antiquark the Casimir operator average is $C_{\bar{Q}}
= 4/3$. Then for the light-heavy system, in a colour-singlet state with 
$C^{0}_{C} = 0$, the colour average operator is
\begin{equation}
\langle\sum_{i=1}^{4} \frac{\lambda_{i}^{c}}{2} . 
\frac{\lambda_{\bar{Q}}^{c}}{2}\rangle = (C^{0}_{C} - C_C - C_{\bar{Q}})/2 
= -4/3 ,
\end{equation}
from which each quark shares 1/4. Hence
\begin{equation}
\langle\frac{\lambda^{c}_{i}}{2} . \frac{\lambda^{c}_{j}}{2}\rangle 
= -1/3 , \hspace{2cm} i= q, \, j=\bar{Q} .
\end{equation}
 This proves that the colour matrix element is $-1/3$ for all pairs. With the
trial wave function (\ref{wf}) the space part of the confining interaction
can be calculated exactly. Then the total confining energy is
\begin{equation}
\langle V_{\text{conf}} \rangle = \frac{C}{2} \left(6 \langle r_{12}\rangle 
+ 4 \langle r_{45}\rangle \right)  
           = \displaystyle \frac{2C}{\sqrt{\pi}} \left(\frac{3}{\sqrt{a}} + 
\sqrt{\frac{3}{2a} + \frac{5}{2b}} \right) .
\end{equation}
The matrix elements of the spin--flavour operators entering the 
GBE interaction (\ref{VCHI}) have been calculated
by using the fractional parentage technique described in Ref.\cite{book}. 
Accordingly, we first determine the explicit form of the spin and flavour 
states
of a given symmetry $[f]_S$ and $[f]_F$ respectively. Then we rewrite each
wave function as a sum of products of the wave function of the first two and
of the last two particles. These are all either symmetric or antisymmetric
two-body states, so that Eqs. (3.3) of Ref.\cite{GR96} can be straightforwardly
applied. In Table \ref{mass}, we exhibit $\langle V_\chi\rangle$ of 
Eq.~(\ref{VCHI}) for the 
cases of interest. The 
upper index $uu$ or $us$ in the orbital two-body matrix element $V$ 
designates the nature of the quark masses in the product 
$m_i m_j$ in Eq.~(\ref{POINT}). 
With the wave function (\ref{wf}), $V$ can be explicitely calculated as:
\begin{equation}
\eqalign{
V^{ij}_\mu=
&{g^2\over4\,\pi}\,{1\over 12\,m_i\,m_j}\,{4\,a^{3/2}\over \pi^{1/2}}\,\Bigg\{
{\mu^2\over 2\,a}\,\exp\left[-r_0(\mu + a\,r_0)\right]
-{\alpha^3\over 2\exp(\alpha^2\,r_0^2)\,(a+\alpha^2)^{3/2} }
\cr
& \Bigg({4\alpha^2\, r_0\over \sqrt{(a+\alpha^2)}\,\sqrt\pi }
 +
\left(2+{4\,\alpha^4\,r_0^2\over a+\alpha^2}\right)\,
\exp\left[{4\alpha^2\,r_0\over a+\alpha^2}\right] \,
\mathop{\rm Erfc}\left[ {-\alpha^2\,r_0\over\sqrt{(a+\alpha^2)}  }\right]
\Bigg) \cr
&- \exp\left[\mu^2/(4\,a)\right] \,
\mathop{\rm Erfc}\left[ {\mu+2\,a\,r_0\over 2\,\sqrt{a} }\right]\,
\mu^3a^{-3/2}/4  \Bigg\}
}
\end{equation}
In Table \ref{mass}, we present the energy 
$E$ of each
$q^4\bar{Q}$ system under discussion, obtained by minimizing $\langle H 
\rangle$ with respect
to $a$ and $b$. Compared to the lowest threshold mass $E_T$ given in the last 
column of
Table \ref{mass}, one can see that the pentaquark energy 
lies several hundred MeV
above the corresponding threshold, in all cases.
One can also notice that the lowering of $E - E_T$ when replacing the quark
$c$ by $b$ is small, of the order of 10 MeV.

Among the systems studied here, the pentaquarks with $S= -2$ appear to be the
most ``favourable'', i.e., have the lowest $E-E_T$, although the value of $E$
of a $uuds\bar{Q}$ system is lower than that of a $udss\bar{Q}$ system,
due to a larger GBE attraction, as discussed above.
 The presence of two strange quarks, instead of one, has a 
negligible effect. Thus the lowering of $E-E_T$ for $udss\bar{Q}$ is 
essentially due to an increase in $E_T$ when passing from the $N + s\bar{Q}$
 threshold to the $\Lambda + s\bar{Q}$ threshold.

In this first simple application of the chiral constituent quark model
\cite{GR96,GPP96,semrel,GR96+} to the pentaquark spectrum, we did not find 
any deeply bound state. We however believe that the present results are
not completely discouraging for the reasons which follow. The energy of 
strange 
pentaquarks is located above the threshold only roughly 1/2 from the amount 
found for hexaquarks with one heavy flavoured quark \cite{hexa}. The values
found for $E$ depend on the approximations involved. A lowering of $E$ can
be achieved by an increase of the basis vectors. A possibility would be to 
consider orbital states of type $[22]_O$ containing two quanta of excitation,
 like in the nucleon-nucleon case \cite{NN97}. The lowest allowed value of
$V_\chi$, Eq.(\ref{opFS}), is $-16\, C_\chi$ as for $|[4]_O [22]_S [211]_F
\rangle$ of Table \ref{states}.
Therefore an interference of the $[4]_O$ state considered here with the
$[22]_O$ state is possible.

Moreover the quark-antiquark interaction was entirely neglected here. 
Assuming that the instanton-induced interaction is very important for 
mesons (see e.g. \cite{Kl90}), one could expect that an additional attractive 
contribution of this type would contribute towards stability of pentaquarks
with respect to the meson--baryon threshold. As mentioned in the 
introduction, the GBE interaction through its Yukawa potential tail can
also generate a medium- and long-range interaction between clusters.
The structure of Eq.~(\ref{POINT}) suggests that, in a frozen configuration,
a strong repulsion at short distances implies a long-range attraction.
Hence our conclusion on the non-existence of deeply-bound compact
pentaquarks encourages pursuing theoretical investigation of 
deuteron-like baryon-meson systems, based on a dynamical approach of a 
few-body system, as for example the resonating group method.

\begin{table}
\renewcommand{\arraystretch}{1.5}
\parbox{18cm}{\caption[states]{\label{states}  The allowed $|[f]_S[f]_F\rangle$ 
states of colour {\bf 3} for the $q^4$ subsystem of pentaquarks with 
strangeness $0,\,-1,\,-2$ and $-3$. The spin $S$ and isospin $I$ are 
indicated in each case
\cite{note}.
The quantity $\langle V_{\chi} \rangle$ is the corresponding expectation 
value of the operator (\ref{opFS}) in units of $C_\chi$.}}
\begin{tabular}{ccccccc}
$|[f]_S[f]_F\rangle$ & $S$ & \multicolumn{4}{c}{$I$} & $\langle V_{\chi}
\rangle$ \\
               &   & $uudd\bar{Q}$ & $uuds\bar{Q}$ & $udss\bar{Q}$ &
 $usss\bar{Q}$ & \\
\tableline
$|[22]_S[211]_F\rangle$ & 0 & / & 1/2 & 0 & / &-16 \\
$|[31]_S[211]_F\rangle$ & 1 & / & 1/2 & 0 & / &-40/3 \\
$|[31]_S[22]_F\rangle$ & 1 & 0 & 1/2 & 1 & /  &-28/3 \\
$|[22]_S[31]_F\rangle$ & 0 & 1 & 1/2,3/2 & 0,1 & 1/2 & -8 \\
$|[4]_S[31]_F\rangle$ & 2 & 1 & 1/2,3/2 & 0,1 & 1/2 & 0 \\
$|[31]_S[4]_F\rangle$ & 1 & 2 & 3/2 & 1 & 1/2 & 8/3 \\
\end{tabular}
\end{table}

\begin{table}
\parbox{18cm}{\caption[hadrons]{\label{hadron} Masses of hadrons required to
calculate the threshold energy $E_T =$ baryon + meson. For $D$ and $B$
 we indicate
$M = (M+3 M^*)/4$. For the baryons, a comparison is shown with the Fadeev
calculations of Ref.~\cite{semrel}}}
\begin{tabular}{c|ccc}
Hadron & \multicolumn{3}{c}{Mass (GeV)} \\
  & variational & Ref.\cite{semrel} & experiment \\
\tableline
N & 0.970 & 0.939 & 0.939 \\
$\Lambda$ & 1.165 & 1.136 & 1.116 \\
$\Sigma$ & 1.235 & 1.180 & 1.192 \\
$\Xi$ & 1.377 & 1.348 & 1.318 \\ 
\tableline
$D$      & 2.008 & & 1.973 \\
$D_s$ & 2.087 &  & 2.076 \\
$B$ & 5.302 & & 5.313 \\ 
$B_s$ & 5.379 & & 5.375
\end{tabular}
\end{table}

\begin{table}
\renewcommand{\arraystretch}{1.5}
\parbox{18cm}{\caption[masses]{\label{mass} Energies of pentaquarks with 
$J=1/2$ and $S=0,\,-1,\,-2$ and $-3$. In column 3, $V_{\gamma}^{ab}$
 $(\gamma = \pi, \eta, K$ or $\eta')$ designates the quark-quark matrix 
elements of the interaction (\ref{POINT}).}}
\begin{tabular}{ccc|c|c|c}
Pentaquark & $I$ & $\langle V_\chi \rangle$ & Energy & 
\multicolumn{2}{c}{Lowest threshold} \\
           &   & (Eq.(\ref{VCHI}))    & (GeV) & $B + M$ & $E_T$ (GeV) \\
\tableline
$\begin{array}{c} uudd\bar{c} \\ uudd\bar{b} \end{array}$ &  $ 
\begin{array}{c}
 0 \\ 0 \end{array} $  & $10 V^{uu}_{\pi} - 2/3 V^{uu}_{\eta} - 4/3
 V^{uu}_{\eta'}$ & $\begin{array}{c} 3.607 \\ 6.889 \end{array}$ &  
$\begin{array}{c}
 N + \bar{D} \\ N + \bar{B} \end{array}$ &  $\begin{array}{c} 2.978 \\
6.272 \end{array}$ \\   
$\begin{array}{c} uuds\bar{c} \\ uuds\bar{b} \end{array}$ & $
\begin{array}{c}
 1/2 \\ 1/2 \end{array} $ & \renewcommand{\arraystretch}{1}
$\begin{array}{c} 7 V^{uu}_{\pi} - 7/9 
V^{uu}_{\eta} - 14/9 V^{uu}_{\eta'} + \\ 22/3 V^{us}_{K} + 22/9 
V^{us}_{\eta} - 22/9 V^{us}_{\eta'} \end{array}$ 
\renewcommand{\arraystretch}{1.5}
 & $\begin{array}{c}  3.545 \\ 6.827 \end{array}$ &  $\begin{array}{c}
 N + \bar{D}_s \\ N + \bar{B}_s \end{array}$ &  $\begin{array}{c} 3.057 \\
 6.349  \end{array}$ \\ 
$\begin{array}{c} udss\bar{c} \\ udss\bar{b} \end{array}$ & $
\begin{array}{c}
 0 \\ 0 \end{array}   $ & \renewcommand{\arraystretch}{1} 
$\begin{array}{c} 4 V^{uu}_{\pi} - 4/9 V^{uu}_{\eta}
 - 8/9 V^{uu}_{\eta'} + \\ 28/3 V^{us}_{K} + 28/9 V^{us}_{\eta} - 
28/9 V^{us}_{\eta'} \end{array}$ \renewcommand{\arraystretch}{1.5}
 & $\begin{array}{c}  3.630 \\ 6.911 \end{array}$ &  $\begin{array}{c}
 \Lambda + \bar{D}_s \\ \Lambda + \bar{B}_s \end{array}$ &  $\begin{array}{c} 
3.253 \\ 6.544
  \end{array}$ \\ 
$\begin{array}{c} usss\bar{c} \\ usss\bar{b} \end{array}$ & $
\begin{array}{c}
 1/2 \\ 1/2 \end{array}   $ & \renewcommand{\arraystretch}{1} 
$\begin{array}{c} 22/3 V^{us}_{K} + 26/9 V^{us}_{\eta} - 
26/9 V^{us}_{\eta'}- \\ 20/9 V^{ss}_{\eta} - 10/9 V^{ss}_{\eta'} \end{array}$ 
\renewcommand{\arraystretch}{1.5}
 & $\begin{array}{c}  3.940 \\ 7.223 \end{array}$ &  $\begin{array}{c}
 \Xi + \bar{D}_s \\ \Xi + \bar{B}_s \end{array}$ &  $\begin{array}{c} 
3.464 \\ 6.756
  \end{array}$ \\ 
\end{tabular}
\end{table}

\begin{references}
\bibitem{Ja77} R.L. Jaffe, Phys. Rev. Lett. {\bf 38} (1977) 195 ; {\bf
38} (1977) 1617(E).
\bibitem{Man95} Proc. 6th. Int. Conf. on Hadron Spectroscopy, Manchester, UK,
July 1995, eds. M.C. Birse, G.D. Lafferty and J.A. McGovern (World Scientific,
Singapore, 1996).
\bibitem{St97} R.W. Stotzer et al., Phys. Rev. Lett. {\bf 78} (1997) 3646.
\bibitem{JMR94} J.-M. Richard, Phys. Rev. {\bf A49} (1994) 3575
\bibitem{MW93} A. V. Manohar and M. B. Wise, Nucl. Phys. {\bf B399} (1993) 17.
\bibitem{Zou86} S. Zouzou, B. Silvestre-Brac, C. Gignoux and J.-M. Richard,
   Z. Phys. {\bf C30}, (1986) 457.
\bibitem{Mo96} M.A. Moinester, Z. Phys. {\bf A355} (1996) 349.
\bibitem{Ka95} D.M. Kaplan, Proc. Int. Workshop Production and Decay of
Hyperons, Charm and Beauty Hadrons, Strasbourg (France), Sept. 5-8, 1995.
\bibitem{Ba96} G. Baum et al., COMPASS Collaboration, CERN-SPSLC-96-14,
March 1996.
\bibitem{Ai97} E.M. Aitala et al., Fermilab E791 Collaboration, 
preprint hep-ex/9709013.
\bibitem{Gi87} C. Gignoux et al., Phys. Lett. {\bf B193} (1987) 323.
\bibitem{Li87} H.J. Lipkin, Phys. Lett. {\bf B195} (1987) 484.
\bibitem{Fl89} S. Fleck et al., Phys. Lett. {\bf B220} (1989) 616.
\bibitem{ZR94} S. Zouzou and J.-M. Richard, Few-Body Syst. {\bf 16} (1994) 1. 
\bibitem{LSB89} J. Leandri and B. Silvestre-Brac, Phys. Rev. {\bf D40} (1989)
2340.
\bibitem{TNK93} S. Takeuchi, S. Nussinov and K. Kubodera, Phys. Lett. {\bf 
B318} (1993) 1.
\bibitem{RS93} D.O. Riska and N.N. Scoccola, Phys. Lett. {\bf B299} (1993) 
338; 
D.P. Min, Phys. Rev. {\bf D50} (1994) 3350; Y. Oh, B.Y. Park and D.P. Min, 
Phys. Lett. {\bf B331} (1994) 36; C.K. Chow, Phys. Rev. {\bf D51} (1995) 6327;
ibidem {\bf D53} (1996) 5108.
\bibitem{Sh95} M. Shmatikov, Phys. Lett. {\bf B349} (1995) 411; Nucl. Phys. 
{\bf A612} (1997) 449. 
\bibitem{charmlets}  J. Schaffner-Bielich and A. P. Vischer, preprint
nucl-th/9710064. 
\bibitem{PSGR97} S. Pepin, Fl. Stancu, M. Genovese and J.-M. Richard,
      Phys. Lett. {\bf B393} (1997) 119.
\bibitem{GR96} L.Ya. Glozman and D.O. Riska, Phys. Rep. 268 (1996) 263.
\bibitem{GPP96} L.Ya. Glozman, Z. Papp and W. Plessas, Phys. Lett. {\bf B381}
(1996) 311.
\bibitem{semrel} L.Ya. Glozman, Z. Papp, W. Plessas, K. Varga and 
R.F. Wagenbrunn, Nucl. Phys. {\bf A623} (1997) 90c.
\bibitem{GR96+} L.Ya. Glozman and D.O. Riska, Nucl. Phys. {\bf A603} (1996)
326.
\bibitem{book} Fl. Stancu, {\it Group Theory in Subnuclear Physics},
Clarendon Press, Oxford (1996).
\bibitem{hexa} Fl. Stancu and S. Pepin, preprint hep-ph/9710528.
\bibitem{NN97} Fl. Stancu, S. Pepin and L. Glozman, Phys. Rev. {\bf C56 } 
 (1997) 2779.
\bibitem{Kl90} C.G. Callan, R. Dashen and D.J. Gross, Phys. Rev. {\bf D17}
(1978) 2717; E. V. Shuryak, Nucl. Phys. {\bf B203} (1982) 93; D.I. Diakonov
and V.Yu. Petrov, Nucl. Phys. {\bf B272} (1986) 457. 
\bibitem{note} The values of the isospin $I$ associated to a given 
hypercharge can be checked
with Table 8.4 of Ref.~\cite{book}.
\end{references}
\end{document}